\documentclass[aps,pra,eqsecnum,twocolumn,showpacs]{revtex4}
\usepackage{epsfig}

\begin{document}
\title{Optimal unambiguous discrimination between subsets of
non-orthogonal quantum states}
\author{Yuqing Sun$^{1}$}
\author{J\'anos A. Bergou$^{1,2}$}
\author{Mark Hillery$^{1}$}
\affiliation{$^1$Department of Physics, Hunter College, City
University of New York, 695 Park Avenue, New York, NY 10021,
USA}
\affiliation{$^{2}$Institute of Physics, Janus Pannonius University,
H-7624 P\'{e}cs, Ifj\'{u}s\'{a}g \'{u}tja 6, Hungary}
\date{\today}
\begin{abstract}
It is known that unambiguous discrimination among non-orthogonal but
linearly 
independent quantum states is possible with a certain probability of
success. Here, we consider a variant of that problem. Instead of
discriminating among all of the different states, we shall only
discriminate between two subsets of them.
In particular, for the case of three non-orthogonal states,
$\{|\psi_1 \rangle, |\psi_2 \rangle,|\psi_3 \rangle \}$, we
show that the optimal strategy to distinguish
$|\psi_1 \rangle$ from the set $\{|\psi_2 \rangle,|\psi_3 \rangle \}$
has a higher success rate than if we wish to discriminate among all
three states. Somewhat surprisingly, for unambiguous discrimination
the subsets need not be linearly independent. A fully analytical
solution is
presented, and we also show how to construct generalized
interferometers (multiports)
which provide an optical implementation of the optimal strategy.
\end{abstract}
\maketitle

\section{Introduction}
According to the quantum theory of measurement, it is impossible to
unambiguously discriminate between
non-orthogonal quantum states with unit success probability. If,
however, we settle for less and don't require that we succeed every
time, then unambiguous discrimination becomes possible.
This procedure uses a non-unitary operation that maps the
non-orthogonal states onto orthogonal ones, and these can then be
discriminated without error using a standard von Neuman measurement.
Although such an operation will always have a certain probability of
failure, we can always tell whether or not the desired transformation
has succeeded. This allows us to achieve unambiguous discrimination.
When the attempt fails, we obtain an inconclusive answer. The optimal
strategy for accomplishing this is the one that minimizes
the average probability of failure.

The problem of unambiguously distinguishing between two non-orthogonal
states was first 
considered by Ivanovic \cite{ivanovic}, and then subsequently by
Dieks \cite{dieks} and Peres \cite{peres2}. These authors found
the optimal solution when the two states are being selected
from an ensemble in which they are equally likely. The optimal
solution for the situation in which the states have different
weights was found by Jaeger and Shimony \cite{jaeger}. We proposed an
optical implementation of the optimal procedure along with a more
compact rederivation of the general results and also showed that the
method is useful in other areas of quantum information processing
\cite{sun1} such
as, for example, entanglement enhancement \cite{chefles4}.
State discrimination measurements have been performed in laboratory,
first by Huttner, et.\ al.\ \cite{huttner} and, more recently, by
Clarke, {\it{et al.}} \cite{clarke}. Both used the polarization
states of photons to represent qubits.
The case of three states was examined by Peres and Terno
\cite{peres3}. It was subsequently extended to the general problem of
discriminating among $N$ states.
Chefles \cite{chefles2} found that $N$ non-orthogonal states
can be probabilistically discriminated without error if and only
if they are linearly independent. Chefles and Barnett \cite{chefles3}
solved the case in which the probability of the procedure succeeding
is the same for each of the states. Duan and Guo \cite{duan}
considered general unitary transformations and measurements on a
Hilbert space containing the states to be distinguished and an
ancilla, which would allow one to discriminate among $N$ states, and
derived matrix inequalities which must be satisfied for the
desired transformations to exist. In our previous paper \cite{sun2},
we presented the necessary conditions for optimal unambiguous
discrimination and used them to derive a method for implementing
the optimal solution. For the case of three states, we presented
optical networks that accomplish this.
One can also consider what happens if the discrimination is not
completely unambiguous, i.\ e.\ if it is possible for errors
to occur, and this was done by Chefles and Barnett \cite{chefles1}.
For an overview of the state-of-the-art on state discrimination see
the excellent recent review article by Chefles \cite{chefrev}.

In these works discrimination among all of the states
was considered. In the present paper, we consider a variant of that
problem. Instead of discriminating among all states, we
ask what happens if we just want to discriminate between subsets of 
them.  A motivation to consider this variant comes from its 
application to comparing strings of qubits in order to find out if
they are identical or not which is certainly one of the basic tasks in
quantum information processing. In particular, if there are three
non-orthogonal states, $\{|\psi_1 \rangle, |\psi_2 \rangle,|\psi_3
\rangle \}$, we wish to find the optimal strategy to unambiguously
distinguish $|\psi_1 \rangle$ from the set $\{|\psi_2 \rangle,|\psi_3
\rangle\}$. We refer to this problem as unambiguous quantum state
filtering. In this context we should note that recently an analytical
solution has been found to the following closely related
problem. Instead of unambiguously distinguishing between two
complementary subsets of an arbitary number $N$ of non-orthogonal
quantum states, occupying a two-dimensional Hilbert space,
errors are allowed but the probability of erroneously assigning the
state to one of the substes is minimized \cite{HB}. The term ``quantum
state filtering'' has been introduced there for the case when one of
the subsets contains one state and the other contains all of the
remaining $N-1$ states. Here, we shall present the analytical solution
for the case of the other possible discrimnation strategy, namely that
of unambiguous  quantum state filtering. 

The paper is divided into six sections. In Section II, based on 
simple but rigorous arguments, we present the 
optimal analytical solution to the problem.
In Section III, we compare these optimal failure probabilities
for two different procedures: discrimination between
$|\psi_1 \rangle$ and $\{|\psi_2 \rangle,|\psi_3 \rangle \}$
and discrimination among all three states. We find that the
failure probability for the first procedure is smaller than
that for the second. In Section IV, we propose a possible experimental 
implementation using the method
proposed in our previous paper \cite{sun2}, which uses a
single-photon representation of the quantum states and an optical
multiport together with photon detection at the output ports to
implement the procedure. A brief discussion and conclusions are given 
in Section V. Finally, in the Appendix, we present an alternative 
derivation, based on the method of Lagrange multipliers, to obtain 
the results of Section II. The method closely parallels the 
techniques used for unambiguous discrimination between all states.

\section{Derivation of the optimal solution}

Suppose we are given a quantum system prepared in the state
$|\psi\rangle$, which is guaranteed to be a member of the set
of three non-orthogonal states $\left\{
|\psi_{1}\rangle,|\psi_{2}\rangle,|\psi_{3}\rangle\right\}$,
but we do not know which one. We want to find a procedure which
will tell us that $|\psi\rangle$ was prepared in
$|\psi_1\rangle$, or will tell us that $|\psi\rangle$ was prepared in
one of $\left\{
|\psi_{2}\rangle,|\psi_{3}\rangle \right\}$. That is, the
procedure can distinguish $|\psi_1\rangle$ from
$\left\{
|\psi_{2}\rangle,|\psi_{3}\rangle \right\}$. We also want this
procedure to be error-free, i.\ e.\ the procedure
may fail to give us any information about the state, and if it
fails, it must let us know that it has, but if it succeeds, it
should never give us a wrong answer. We shall refer to such a
procedure as quantum state filtering without error. We find that, in 
contrast to the unambiguous state discrimination problem, this will
be possible even if $|\psi_1 \rangle$ is not linearly independent
from the set $\{|\psi_2 \rangle,|\psi_3 \rangle\}$.

If the states are not orthogonal then, according to the quantum
theory of measurement, they cannot be discriminated perfectly.
In other words, if we are given $|\psi_{i}\rangle$, we will have
some probability $p_{i}$ to determine what it is
successfully and, correspondingly, some
failure probability, $q_{i}= 1-p_{i}$, to obtain an inconclusive
answer.
If we denote by $\eta_i$ the {\em a priori} probability that the
system was
prepared in the state $|\psi_{i}\rangle$, the average probabilities
of success and of failure to distinguish the states $|\psi_{i}\rangle$
are
\begin{eqnarray}
\label{Psf}
P &=& \sum_{i}{\eta}_{i}p_{i} , \nonumber \\
Q &=& \sum_{i}{\eta}_{i}q_{i} ,
\end{eqnarray}
respectively.
Our objective is to find the set of $\left\{ p_i \right\}$ that
maximizes the probability of success, $P$.

The procedure we shall use is a ``generalized measurement'', which can
be described as follows. Let $\mathcal{K}$ denote a total Hilbert 
space,
which is the direct sum
of two subspaces, $\mathcal{K}=\mathcal{H}\oplus\mathcal{A}$.
The 
space $\mathcal{H}$ is a three-dimensional space that contains the
vectors $|\psi_{i}\rangle$, and $\mathcal{A}$ is an auxiliary space.
The input state of the system is one of the vectors 
$|\psi_{i}\rangle$,
which is now a vector in the subspace $\mathcal{H}$ of the total
space $\mathcal{K}$, so that
\begin{equation}
|\psi_i^{\mathcal K} \rangle_{in} =
|\psi_i^{\mathcal H} \rangle .
\label{instate}
\end{equation}
A unitary transformation, $U$, which acts in
the entire space $\mathcal{K}$ is now applied to the input vector,
resulting in the state $|\psi_i^{\mathcal K} \rangle_{out}$, which is
given by
\begin{equation}
\label{outstate}
|\psi_i^{\mathcal K} \rangle_{out} =
|\psi_i^{\prime \, \mathcal H} \rangle + |\phi_i^{\mathcal A} \rangle 
=
U |\psi_i^{\mathcal K} \rangle_{in} ,
\end{equation}
where, in our case, $| \psi_1^{\prime} \rangle$ can always be
unambiguously distinguished from the set $\{|\psi_2^{\prime}
\rangle,|\psi_3^{\prime} \rangle \}$. Then a measurement
is performed on $|\psi_i^{\mathcal K} \rangle_{out}$ that
projects $|\psi_i^{\mathcal K} \rangle_{out}$ either onto
$|\psi_i^{\prime} \rangle$ or $| \phi_i \rangle$ (by construction,
they are in orthogonal subspaces). If it projects
$|\psi_i^{\mathcal K} \rangle_{out}$ onto $| \psi_i^{\prime} \rangle$,
the procedure succeeds, because $|\psi_1^{\prime} \rangle$ can always
be distinguished from $\{|\psi_2^{\prime} \rangle,|\psi_3^{\prime}
\rangle \}$. The probability to get this outcome, if the input state
is $|\psi_{i}\rangle$, is
\begin{equation}
\label{pi}
p_i= \langle \psi_i^{\prime}|\psi_i^{\prime} \rangle.
\end{equation}
If the measurement
projects
$|\psi_i^{\mathcal K} \rangle_{out}$ onto $| \phi_i \rangle$, the
procedure fails.
The probability of this outcome is
\begin{equation}
\label{q_i}
q_i= 1-p_i=\langle \phi_i|\phi_i \rangle.
\end{equation}

The nature of the problem we are trying to solve imposes a number
of requirements on the output vectors.
The condition that $|\psi_1^{\prime} \rangle$ be
distinguishable from $|\psi_2^{\prime} \rangle$ and
$|\psi_3^{\prime}\rangle$ requires that
\begin{equation}
\label{orthogonal}
\langle \psi_1^\prime|\psi_2^\prime \rangle =
\langle \psi_1^\prime|\psi_3^\prime \rangle =0.
\end{equation}
These lead to conditions on the failure vectors, $|\phi_{i}\rangle$.
Taking the scalar product between $|\psi^{\mathcal K}_{1}
\rangle_{out}$ and the other two output states
and using Eq.\ (\ref{orthogonal}) and the fact that $U$
is unitary leads to the conditions
\begin{eqnarray}
\label{phi_1213}
\langle \phi_1|\phi_2 \rangle &=& \langle \psi_1|\psi_2 \rangle,
\nonumber \\
\langle \phi_1|\phi_3 \rangle &=& \langle \psi_1|\psi_3 \rangle.
\end{eqnarray}
Our objective is to find the optimal $| \psi_i^{\prime} \rangle$
and $|\phi_i \rangle$
which satisfy Eqs.\ (\ref{pi})--\ (\ref{phi_1213}) and also give the
maximum success probability $P$.

Let us now consider the failure vectors. If they were linearly
independent, we could apply a state discrimination procedure to
them \cite{chefles2}.
That means that if our original procedure fails, and
we end up in the failure space, $\mathcal{A}$, then we still
have some chance of determining what our input state was. This
clearly implies that our original procedure, which led to the
vectors $|\psi^{\prime}\rangle$, was not optimal, because that
process followed by another on the failure vectors would lead to
a higher probability of distinguishing $|\psi_{1}\rangle$
from $|\psi_{2}\rangle$ and $|\psi_{3}\rangle$. Therefore,
the optimal procedure should lead to failure vectors to
which we cannot successfully apply a state discrimination
procedure, implying that they are linearly dependent. In fact, we will 
now prove that for optimal discrimination they must be collinear, by 
demonstrating that the contrary leads to contradiction. To this end, 
we assume that we have achieved optimal unambiguous discrimination of 
$|\psi_1 \rangle$ from $|\psi_2\rangle$ and $|\psi_3\rangle$ 
but the failure vectors are {\it not} collinear. Then at least one of 
the two failure vectors, $|\phi_2 \rangle$, $|\phi_3 \rangle$, will 
have a component in the direction that is perpendicular to 
$|\phi_{1}\rangle$. We can set up a detector projecting onto this 
direction and a positive outcome of the measurement (a click of the 
detector) will tell us that our input state was not $|\psi_{1}\rangle$ 
but one of the other two states. Thus, contrary to our assumption that 
our procedure has been optimal, further distinction is
possible. Hence, the  failure vectors must be collinear for optimal
discrimination.    

We shall now explore the consequences of this conclusion.
Since $|\phi_i \rangle$ ($i=1,\ldots, n)$ are collinear,
the failure space, $\mathcal{A}$, is one dimensional. If 
$|u\rangle$ is the basis vector spanning this Hilbert space we can 
write the failure vectors as $|\phi_i 
\rangle=\sqrt{q_i}e^{\chi_i}|u\rangle$. Substituting this 
representation of the failure vectors into Eq.
(\ref{phi_1213}), we find that
\begin{eqnarray}
    \label{TwoDeltas}
q_1q_2 &=& | \langle \psi_1|\psi_2 \rangle|^2,
\nonumber \\
q_1q_3 &=& | \langle \psi_1|\psi_3 \rangle|^2.
\end{eqnarray} 
These two conditions are a consequence of unitarity and imply that 
only one of the three failure probabilities can be chosen 
independently. If we chose $q_1$ as the independent one we can 
express the other two as
$q_2 = | \langle \psi_1|\psi_2 \rangle |^2/q_1$ and
$q_3 = | \langle \psi_1|\psi_3 \rangle |^2/q_1$.
If we introduce the notation $O_{ij}= \langle \psi_i|\psi_j \rangle$ 
then, with the help of these two equations, the average failure 
probability can be written explicitly as
\begin{eqnarray}
Q &=& \sum_i \eta_{i} q_i
\nonumber
\\
&=& \eta_1 q_1 +  \frac{\eta_2 |O_{12}|^2 + \eta_3 |O_{13}|^2}{q_1 }.
\end{eqnarray}
If we further introduce the notation $A=\eta_2 |O_{12}|^2 + \eta_3 
|O_{13}|^2$ for the frequently occuring average overlap 
then, from the condition 
\begin{equation}
\frac{d Q}{d q_1}=0,
\end{equation}
we find the optimal value of $q_1$ to be
\begin{equation}
    q_{1}=\sqrt{A/ \eta_1} .
    \label{q1opt}
\end{equation}

This value, however, cannot always be realized. 
For it to be true, there must be a unitary transformation, from Eq. 
(\ref{outstate}), that takes $|\psi_{j}\rangle$
to $|\psi_{j}\rangle_{out}$ which, together with the one-dimensionality 
of the failure space yields
\begin{equation}
|\psi_{j}\rangle_{out}=|\psi_{j}^{\prime}\rangle +\sqrt{q_{j}}
|e^{i\chi_{j}} |u\rangle .
\end{equation}
Here we have that $\langle\psi_{j}^{\prime}|u\rangle =0$, $\langle
\psi_{1}^{\prime}|\psi_{j}^{\prime}\rangle =0$ for $j=2,3$, and 
the phase factors are fixed by the requirement (cf. Eq. 
(\ref{phi_1213})) that
\begin{equation}
\langle\psi_{1}|\psi_{j}\rangle = 
\sqrt{q_{1}q_{j}}e^{i(\chi_{j}-\chi_{1})} \,
\end{equation}
for $j=2,3$.  These equations imply that
\begin{equation}
\langle\psi_{j}^{\prime}|\psi_{k}^{\prime}\rangle = \langle\psi_{j}|
\psi_{k}\rangle - \sqrt{q_{j}q_{k}}e^{i(\chi_{k}-\chi_{j})} .
\end{equation}
This set of equations can only be true if the matrix $M$, where
\begin{equation}
M_{jk}= \langle\psi_{j}|
\psi_{k}\rangle - \sqrt{q_{j}q_{k}}e^{i(\chi_{k}-\chi_{j})} ,
\end{equation}
is positive semidefinite, as discussed in detail in Ref. \cite{sun2}.

Using again $O_{jk}=\langle\psi_{j}|\psi_{k}\rangle$, $M$ can be 
expressed as
\begin{equation}
M=\left( \begin{array}{ccc} 1-q_{1} & 0 & 0 \\ 0 & 1-\frac{|O_{12}|^{2}}
{q_{1}} & O_{23}-\frac{O_{21}O_{13}}{q_{1}} \\ 0 & O_{32}
-\frac{O_{31}O_{12}}{q_{1}} & 1-\frac{|O_{13}|^{2}}{q_{1}}
\end{array}\right) .
\end{equation}
Clearly, this matrix will be positive semidefinite if $0\leq q_{1}\leq 1$, 
and if the $2\times 2$ submatrix is also positive semidefinite.  This
will be true if both the trace and determinant of the submatrix
are greater than or equal to zero.  Positivity requires that the diagonal
matrix elements of the submatrix be non-negative, so that it must be
true that $q_{1}\geq |O_{12}|$ and $q_{1}\geq |O_{13}|$ .  Without loss 
of generality, we can assume that $|O_{12}|\geq |O_{13}|$ by simply 
arranging the states in set $2$ in the order of decreasing 
overlaps with $|\psi_{1}\rangle$.  Doing so
and imposing the condition that  $q_{1}\geq |O_{12}|$ guarantees that
the condition $q_1\geq|O_{13}|$ is also satisfied, and together they
imply that the trace is greater than or equal to zero.

The condition that the determinant be non-negative gives us a lower 
bound on $q_{1}$,
\begin{equation}
\label{q1ineq}
q_{1}\geq \frac{|O_{12}|^{2}+|O_{13}|^{2}-(O_{12}O_{23}O_{31}
+O_{13}O_{32}O_{21})}{1-|O_{23}|^{2}} .
\end{equation}
We want to interpret this inequality, in particular, we want to find what
the right-hand side is equal to.  In order to do so, we shall find the
projection operator, $P_{23}$, that projects onto the subspace spanned by 
$\psi_{2}$ and $\psi_{3}$. One of the basis vectors in this subspace 
can be chosen to be $|\psi_{2}\rangle$ and, using the Gram-Schmidt 
orthogonalization method, the other is defined as the (normalized) 
orthogonal component of $|\psi_{3}\rangle$,
\begin{equation}
|{\tilde\psi}_{3}\rangle = \frac{1}{\sqrt{1-|O_{23}|^{2}}}
(|\psi_{3}\rangle -O_{23}|\psi_{2}\rangle ) .
\label{basis3}
\end{equation}
This leads to
\begin{equation}
P_{23}=|\psi_{2}\rangle\langle\psi_{2}| + |{\tilde\psi}_{3}\rangle
\langle{\tilde\psi}_{3}| .
\label{projector23}
\end{equation}
Let us represent the input state, $|\psi_{1}\rangle$, as
$|\psi_{1}\rangle = 
|\psi^{\perp}_{1}\rangle + |\psi^{\parallel}_{1}\rangle$, where 
$|\psi^{\perp}_{1}\rangle=(1-P_{23})|\psi_{1}\rangle$ is the component
of the input vector that is perpendicular to the subspace spanned by 
$\psi_{2}$ and $\psi_{3}$ and $|\psi^{\parallel}_{1}\rangle =
P_{23}|\psi_{1}\rangle$ is the 
component in that subspace. Then, using Eqs. (\ref{basis3}) and
(\ref{projector23}), the explicit expression for the parallel
component is given by
\begin{equation}
|\psi^{\parallel}_{1}\rangle =
\frac{O_{21}-O_{23}O_{31}}{1-|O_{23}|^{2}} |\psi_{2}\rangle +
\frac{O_{31}-O_{32}O_{21}}{1-|O_{23}|^{2}} |\psi_{3}\rangle].
\label{psiparallel}
\end{equation}
Calculating the norm of this expression yields
\begin{equation}
\langle\psi^{\parallel}_{1}|\psi^{\parallel}_{1}\rangle = 
 \frac{|O_{12}|^{2}+|O_{13}|^{2}-(O_{12}O_{23}O_{31}
+O_{13}O_{32}O_{21})}{1-|O_{23}|^{2}} ,
\label{psiparallelnorm}
\end{equation} 
which is identical to the right-hand side of Eq.\
(\ref{q1ineq}).

Thus, Eq. (\ref{q1ineq}) tells us that the failure
probability, $q_{1}$, has a lower bound which is given by the weight
of $|\psi_{1}\rangle$ in the other subspace,
$\| P_{23}\psi_{1}\|^{2}=\langle\psi_{1}|P_{23}|\psi_{1}\rangle = 
\langle\psi^{\parallel}_{1}|\psi^{\parallel}_{1}\rangle$, a result
that is intuitively obvious. Clearly, 
this expression is larger than (or at most equal to) $|O_{12}|^{2}$.
This implies that, because $q_{2}=|O_{12}|^{2}/q_{1}$, we have
\begin{eqnarray}
q_{2}& \leq & \frac{|O_{12}|^{2}} 
{\langle\psi^{\parallel}_{1}|\psi^{\parallel}_{1}\rangle} 
\nonumber \\
 & = & \frac{|O_{12}|^{2}} {|O_{12}|^{2}+|\langle{\tilde\psi}_{3}|
\psi_{1}\rangle |^{2}} \leq 1 ,
\end{eqnarray}
and similarly for $q_{3}$. 

We can then distinguish three 
different regimes of the parameters. If the r.h.s. of Eq.\ 
(\ref{q1opt}) is greater than $1$ then $q_{1}=1$, if it is less 
than $\langle\psi^{\parallel}_{1}|\psi^{\parallel}_{1}\rangle$ then 
$q_{1}=\langle\psi^{\parallel}_{1}|\psi^{\parallel}_{1}\rangle$, 
and in the intermediate range the optimum given by Eq. (\ref{q1opt}) 
is realized. This can be summarized as follows.

(i) If $\eta_{1} 
|\langle\psi^{\parallel}_{1}|\psi^{\parallel}_{1}\rangle|^{2} \le A \le 
\eta_{1}$, then
\begin{eqnarray} 
\label{result1} 
q_1 &=& \sqrt{A/\eta_{1}} ,
\nonumber \\ 
q_2 &=& \sqrt{\eta_{1}/A} |O_{12}|^2 ,
\nonumber \\
q_3 &=& \sqrt{\eta_{1}/A} |O_{13}|^2 ,
\end{eqnarray}
yielding the average failure probability
\begin{equation}
    Q = 2 \sqrt{\eta_{1} A} .
    \label{Q1}
\end{equation}

(ii) If $ A \geq \eta_{1} $, then
\begin{eqnarray} 
\label{result2} 
q_1 &=& 1,
\nonumber \\ 
q_2 &=& |O_{12}|^2 ,
\nonumber \\ 
q_3 &=& |O_{13}|^2.
\end{eqnarray}
yielding the average failure probability
\begin{equation}
    Q= \eta_{1} + A .
    \label{Q2}
\end{equation}

(iii) If $ A \le \eta_{1} 
|\langle\psi^{\parallel}_{1}|\psi^{\parallel}_{1}\rangle|^{2} $, then
\begin{eqnarray} 
\label{result3} 
q_1 &=& \langle\psi^{\parallel}_{1}|\psi^{\parallel}_{1}\rangle,
\nonumber \\ 
q_2 &=& \frac{|O_{12}|^{2}} 
{\langle\psi^{\parallel}_{1}|\psi^{\parallel}_{1}\rangle} ,
\nonumber \\ 
q_3 &=& \frac{|O_{13}|^2} 
{\langle\psi^{\parallel}_{1}|\psi^{\parallel}_{1}\rangle} ,
\end{eqnarray}
yielding the average failure probability
\begin{equation}
    Q= \eta_{1} 
    \langle\psi^{\parallel}_{1}|\psi^{\parallel}_{1}\rangle  + 
    \frac{A} 
    {\langle\psi^{\parallel}_{1}|\psi^{\parallel}_{1}\rangle} .
    \label{Q3}
\end{equation}

Equations (\ref{result1})--(\ref{Q3}) summarize our main results. In the 
intermediate range of the average overlap, $A$, the optimal failure 
probability, Eq. (\ref{Q1}), is achieved by a generalized measurement 
or POVM. Outside this region, for very large average overlap, $A \geq 
\eta_{1}$, or very small average overlap, $A \leq \eta_{1} 
|\langle\psi^{\parallel}_{1}|\psi^{\parallel}_{1}\rangle|^{2}$, the 
optimal failure probabilities, Eqs. (\ref{Q2}) and (\ref{Q3}), are 
realized by standard von Neumann measurements. For very large $A$ the 
optimal von Neumann measurement consists of projections onto 
$|\psi_{1}\rangle$ and two orthogonal directions whose 
directionality needs not be specified further. A click along 
$|\psi_{1}\rangle$ corresponds to failure because it can have its 
origin in any of the two subsets and a click in the orthogonal 
directions uniquely assigns the input state to the set 
$\{|\psi_{2}\rangle,|\psi_{3}\rangle\}$. For very small $A$ the 
optimal von Neumann measurement consists of projections onto 
$|\psi^{\parallel}_{1}\rangle$ and two orthogonal directions 
that are uniquely determined by the requirement that they 
correspond to two mutually exclusive alternatives. One of them is onto 
$|\psi^{\perp}_{1}\rangle$ and the other onto the remaining orthogonal 
direction in the subspace of $\{|\psi_{2}\rangle,|\psi_{3}\rangle\}$. 
A click along $|\psi^{\parallel}_{1}\rangle$ corresponds to failure 
because it can originate from any of the input states while a click in 
any of the alternative directions unambiguously assigns the input to 
one or the other of the two mutually exclusive subsets. It is 
interesting to observe that the failure space is one dimensional for
each of the three different optimal measurements in the three
different regions. At the boundaries of their respective
regions of validity, the optimal measurements transform into one
another continuously. Furthermore, each of the two von Neumann
expressions can be written as the arithmetic 
mean of two terms and the POVM result as the geometric mean of the 
same two terms. Therefore, in its range of validity the POVM performs 
better than any von Neumann measurement.

In closing this Section we
want to point out an interesting feature of the solution. The results
hold true even when there is no perpendicular component of the first
input state, $|\psi^{\perp}_{1}\rangle=0$, i.e. it lies
entirely in the Hilbert space spanned by the other two vectors or, in
other words, the two sets are linearly dependent. In this case the two
von Neumann  measurements coincide and the range of validity of the
POVM solution shrinks to zero. A click in the detector along the first
input vector corresponds to failure - it might originate from either
of the two subsets - and a click in the detector along the single
direction orthogonal to it unambiguously identifies the set of
the other two vectors. 

An alternative derivation of the above results, that is based on 
the method of Langrange multipliers, is given in the Appendix.

\section{Comparison to the case when all states are discriminated}

In this section
we want to compare the average probability of failure $Q$ of the
filtering problem to that of distinguishing all three states.
Let $Q^{\prime}$ denote the average probability of failure
for distinguishing all the states $\{ |\psi_1\rangle,|\psi_2\rangle,
|\psi_3\rangle \}$. We can see immediately,
that the probability of failure to
distinguish $|\psi_1\rangle$ from $\{ |\psi_2\rangle, |\psi_3\rangle
\}$, $Q$, should be no larger than $Q^{\prime}$.
For the latter problem, the necessary condition for achieving optimal
discrimination is
\begin{equation}
\label{delta3state}
\left|
\begin{array}{ccc}
q_1 & O_{12} & O_{13}
\\ 
O_{12}^{\ast} & q_2 & O_{23}
\\
O_{13}^{\ast} & O_{23}^{\ast} & q_3
\end{array} \right|
=0.
\end{equation}
When comparing this equation to Eq.\ (\ref{delta}),
we see that, instead of a given constant $O_{23}$
that appears in Eq.\ (\ref{delta3state}),
there are the variables $r$ and $\theta$ in Eq.\ (\ref{delta}).
These variables are chosen to minimize the average probability of 
failure $Q$. Therefore, $Q$ should be no larger than $Q^{\prime}$, 
$Q \leq Q^{\prime}$. 

To illustrate this point, we use a simple symmetric case,
where all of the overlaps between the states are real and equal,
\begin{equation}
\langle \psi_1|\psi_2 \rangle = \langle \psi_1|\psi_3 \rangle
= \langle \psi_2|\psi_3 \rangle = s,
\end{equation} 
with $0<s<1$. 
We shall also assume that the {\em a priori} probabilities are equal
for all the examples in this paper. From previous work we know that
in this case, the optimal values of the failure probabilities when we
wish to distinguish among all of
the states $\{|\psi_1 \rangle,|\psi_2 \rangle,|\psi_3 \rangle \}$
are $q_i=s$, which implies that $Q^{\prime}=s$ \cite{sun2}.

\begin{figure}[ht]
\epsfig{file=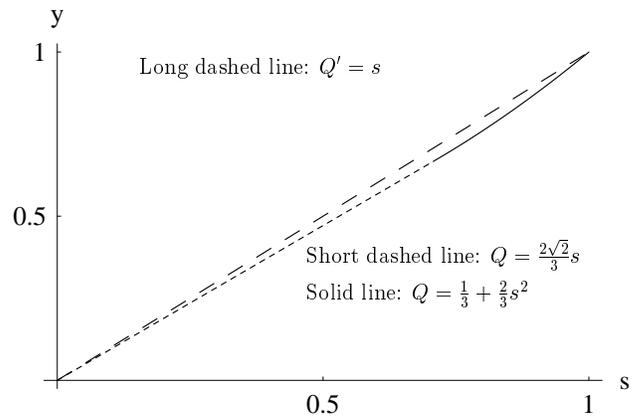, height=5.5cm}
\caption{We compare $Q$ and $Q^{\prime}$. For $0<s \leq
\frac{\sqrt{2}}{2}$ we have that $Q^{\prime}=s$ and
$Q=\frac{2 \sqrt{2}}{3}s$. For $\frac{\sqrt{2}}{2}<s \leq 1$, we still
have that $Q^{\prime}=s$, but $Q = \frac{1}{3} + \frac{2}{3}s^2$.
Note that $Q$ is always smaller than $Q^{\prime}$.}
\label{figSolution}
\end{figure}

For the problem of distinguishing $|\psi_1 \rangle$ from $\{|\psi_2
\rangle,|\psi_3 \rangle \}$,
from the results of Eqs.\ (\ref{result1}) and (\ref{result2}),
we have
(i) if $0< s \leq \frac{\sqrt{2}}{2}$, then
\begin{eqnarray} 
\label{symSolution1}
q_1 = \sqrt{2}s, 
\nonumber \\ 
q_2 = q_3 =\frac{\sqrt{2}}{2}s,
\nonumber \\
Q = \frac{2 \sqrt{2}}{3}s .
\end{eqnarray}
So the average probability of failure $Q$ is less than $Q^{\prime}=s$.
(ii) if $\frac{\sqrt{2}}{2} <s < 1$, then
\begin{eqnarray} 
\label{symSolution2}
q_1 = 1,
\nonumber \\ 
q_2 = q_3 = s^2,
\nonumber \\
Q = \frac{1}{3} + \frac{2}{3}s^2.
\end{eqnarray}
These solutions are illustrated and compared to $Q^{\prime}$ in
Figure \ref{figSolution}. Note that in both cases we have that
$Q < s=Q^{\prime}$.

Now we shall compare filtering to the problem
of distinguishing two states $\{|\psi_1 \rangle,|\psi_2 \rangle \}$,
when all the {\em a priori} probabilities are equal.
If we denote by $Q^{\prime \prime}$ the average probability of failure
when distinguishing between the two states $\{|\psi_1 \rangle$ and
$|\psi_2 \rangle \}$,
we know that $Q^{\prime \prime}=|O_{12}|$ 
(Refs. \cite{ivanovic}--\cite{jaeger}).
For the case we are considering, $|O_{12}|=|O_{13}|=s$, and
we see that $Q < Q^{\prime \prime}$.

A second example is more illuminating. The overlaps are now
given by
\begin{eqnarray}
\langle\psi_{1}|\psi_{2}\rangle &=& \langle\psi_{1}|
\psi_{3}\rangle = s_{1}, \nonumber \\
\langle\psi_{2}|\psi_{3}\rangle &=& s_{2} ,
\end{eqnarray}
where, for simplicity, $s_{1}$ and
$s_{2}$ are real, $0< s_1, s_2 <1$, and
\begin{equation}
0<s_1< \frac{\sqrt{2}}{2}, \text{ } s_1^2<s_2, \text{ and }
s_1<2 s_2. \nonumber
\end{equation}
The probabilities of failure for discriminating
$|\psi_1 \rangle$ from $\{|\psi_2 \rangle,|\psi_3 \rangle \}$ are
\begin{eqnarray} 
q_1 &=& \sqrt{2}s_1,
\nonumber \\ 
q_2 &=& q_3 = \frac{\sqrt{2}}{2}s_1 ,
\label{quneven}
\end{eqnarray}
and the average failure probability is
\begin{eqnarray}
\label{Quneven}
Q = \frac{2 \sqrt{2}}{3}s_1 .
\end{eqnarray}
The optimal probabilities of failure for
discriminating among all three states $\{|\psi_1 \rangle,|\psi_2
\rangle,|\psi_3 \rangle \}$ are given by \cite{sun2}
\begin{eqnarray}
q_{1}^{\prime} &=& \frac{s_{1}^{2}}{s_{2}} \nonumber \\
q_{2}^{\prime} &=& q_{3}^{\prime}= s_{2} \nonumber \\
Q^{\prime} &=& \frac{1}{3} [(s_{1}^{2}/s_{2})+2s_{2}].
\end{eqnarray}

$Q$ can be compared to $Q^{\prime}$ by examining the ratio
\begin{equation}
\frac{Q}{Q^{\prime}} =
\frac{2\sqrt{2} s_1s_2}{s_1^{2} + 2 s_2^{2}} \leq 1 .
\end{equation}
From the above equation, we see that when $s_1$ is much
smaller than $s_2$, $Q$ is much smaller than $Q^{\prime}$.
For example, when $s_1= \frac{\sqrt{2}}{5}$, $s_2=\frac{4}{5}$,
$Q/Q^{\prime} = 0.47$.

\section{Optical realization}

Now we shall present a scheme for a possible experimental realization
of the optimal discrimination
between $|\psi_1 \rangle$ and $\{|\psi_2 \rangle,|\psi_3 \rangle \}$.
The method is similar to one we proposed in a previous publication
\cite{sun2}. We shall use single photon states to represent the input
and output states,
and an optical eight-port together with photon detectors placed at
the output ports to realize the unitary transformation and subsequent
measurements.

Our states will be a single photon split among several modes. Each
mode will serve as an input to an optical eight-port.
Recall that the dimension of the total Hilbert space is four, so
we shall require four modes, and the input
states $|\psi_i \rangle$ will be represented by
single photon states as
\begin{equation}
|\psi_i \rangle =\sum_{j=1}^{4}d_{i j} \hat{a}_{j}^{\dagger} 
|0\rangle,
\end{equation}
where $\sum_{j=1}^{4}|d_{i j}|^2=1$, and $\hat{a}_{j}^{\dagger}$ is 
the
creation operator for the $j$th mode. We shall require $d_{i4}=0$ for
$i=1,2,3$, that is, the initial single photon state is sent
to the first three input ports, and the vacuum into the fourth input
port. The first three modes correspond to the space, $\mathcal{H}$,
containing the states to be distinguished and the fourth mode to
the failure space, $\mathcal{A}$.

\begin{figure}[ht]
\epsfig{file=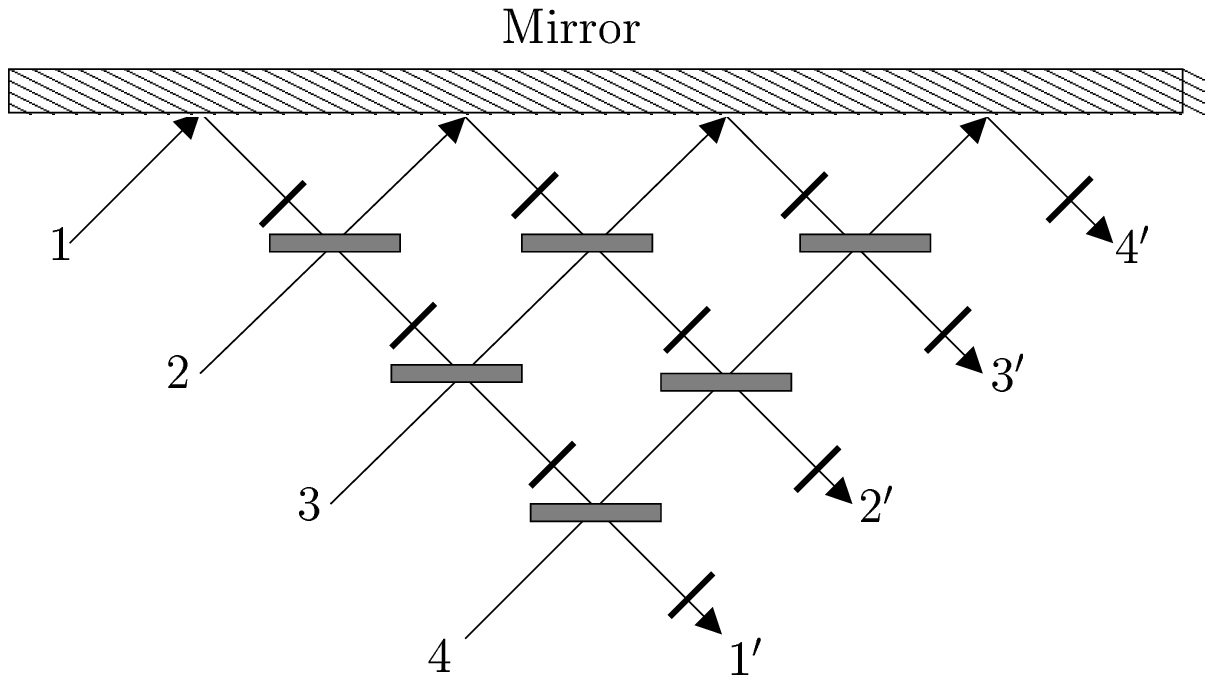, height=5cm}
\caption{An optical eight-port. The beams are straight lines, a
suitable beam splitter is placed at each point where two beams
intersect, 
phase shifters are at one input of each beam splitter and at each 
output.}
\label{etport1}
\end{figure}

In general, an optical $2N$-port is a lossless linear device with $N$
input ports and $N$ output
ports. Its action on the input states can be described by a
unitary operator, $U_{2N}$, and physically it consists of an 
arrangement
of beam splitters, phase shifters, and mirrors.
Since the dimension of the input and output states is four, here we 
shall
use an eight-port (see Figure \ref{etport1}).

\noindent
If we denote the annihilation operators corresponding to the input
modes of the eight-port by
$a_{j}$, $j=1,\ldots,4$, then the output operators are
given by
\begin{equation}
\label{trans}
a_{j\bf{out}}=U^{-1} a_{j}U
=\sum_{k=1}^{4}M_{jk}a_{k},
\end{equation}
where $M_{jk}$ are the elements of a $4\times 4$ unitary
matrix $M(4)$. In the
Schr\"{o}dinger picture, the $in$ and $out$ states are
related by
\begin{equation}
|\psi \rangle_{out} = U |\psi\rangle_{in} .
\end{equation}
It can be shown \cite{sun2} that when using single photon states
representation,
the matrix element
$M_{il}$ is the same as the matrix element of $U $ between the
single-particle states $|i\rangle = a^{\dagger}_{i}|0\rangle$ and
$|l\rangle = a^{\dagger}_{l}|0\rangle$, i.e.,
\begin{equation}
\label{UToM}
\langle i|U |l\rangle = M_{il} .
\end{equation}

To design the desired eight-port, we first calculate the
optimal value of $q_i$. Then from Eq.\ (\ref{q_i}) and the
fact that our failure space is one-dimensional, the vectors
$|\phi_i \rangle$ are given by
\begin{equation}
|\phi_i\rangle=
\sqrt{q_i} |1^{\mathcal A}\rangle = \sqrt{q_i}a^{\dagger}_{4}
|0\rangle ,
\end{equation}
where the state $|1^{\mathcal A}\rangle$ denotes one photon state
in the failure space, which is just one photon in mode $4$.
Once the vectors $|\phi_i \rangle$ are determined, the inner products
$\langle \psi_i^{\prime}|\psi_j^{\prime} \rangle$ $(i,j=1,2,3)$ are 
given
by
\begin{equation}
\label{psi_i'j'}
\langle \psi_i^{\prime}|\psi_j^{\prime} \rangle =
\langle \psi_i|\psi_j \rangle_{in} - \langle \phi_i|\phi_j \rangle.
\end{equation}
We then have to find vectors $|\psi_i^{\prime} \rangle$ that satisfy
this equation. The answer is not unique, and one way of proceeding
is the following. If we define the hermitian matrix $L$ to be
\begin{equation}
L_{ij}=\langle \psi_i|\psi_j \rangle_{in}
- \langle \phi_i|\phi_j \rangle,
\end{equation}
then we note from Eq. (\ref{phi_1213}) that $L_{12}=L_{13}=0$.
This implies that the simplest choice for $|\psi_{1}^{\prime}\rangle$
is a vector with only one nonzero component. Then the vectors
$|\psi_{2}^{\prime}\rangle$ and $|\psi_{3}^{\prime}\rangle$ will have
nonzero components in only their other two places. The obvious choice
is 
\begin{equation}
|\psi_{1}^{\prime}\rangle =\left( \begin{array}{c} \sqrt{p_{1}} \\
0 \\ 0 \\ 0 \end{array}\right) .
\end{equation}
In this column vector, the first entry is the amplitude of the
photon to be in mode $1$, the second is the amplitude to be in
mode $2$, etc. Mode $4$ corresponds to the failure space,
$\mathcal{A}$. The vectors $|\psi_{2}^{\prime}\rangle$ and
$|\psi_{3}^{\prime}\rangle$ will have nonzero components in
only their second and third places, and
if their overlap is real, we can choose
\begin{equation}
|\psi_{2}^{\prime}\rangle =\left( \begin{array}{c} 0 \\ \sqrt{p_{2}}
\cos\theta \\ \sqrt{p_{2}}\sin\theta \\ 0 \end{array}\right) , \
|\psi_{3}^{\prime}\rangle =\left( \begin{array}{c} 0 \\ \sqrt{p_{3}}
\cos\theta \\ -\sqrt{p_{3}}\sin\theta \\ 0 \end{array}\right) ,
\end{equation}
where 
\begin{equation}
\theta = \frac{1}{2}\cos^{-1}\left(\frac{L_{23}}{\sqrt{p_{2}p_{3}}}
\right) .
\end{equation}
This simple choice works for the last example in this section
(see Eq.\ (\ref{output2}), below). For the
first, somewhat more general, example we are forced to choose the
second component of $|\psi_{1}^{\prime}\rangle$ to be nonzero and
then the first and third components of the other two success vectors
are different from zero. They can be obtained by simply interchanging
the first and second components in the above expressions of the
vectors $|\psi_i^{\prime} \rangle$ (see Eq.\ (\ref{output1}),
below). 

Once we have the input and output vectors, the unitary transformation,
$U$, which maps the input states onto
the output states then can be chosen, and this,
as shown by Eq.\ (\ref{UToM}), gives the explicit form of $M(4)$.
Furthermore, $M(4)$
can be factorized as a product of two-dimensional
$U(2)$ transformations\cite{sun2,reck},
and any $U(2)$ transformations can be implemented by a lossless
beam splitter and a phase shifter with appropriate parameters.
A beam splitter with a
phase shifter at one output port transforms the input operators
into output operators as
\begin{equation}
\label{splitter}
\left( \begin{array}{c}
a_1 \\ a_2
\end{array} \right)_{out} = \left(
\begin{array}{cc}
e^{i \phi} \sin \omega & e^{i \phi} \cos \omega \\
\cos \omega & - \sin \omega
\end{array} \right) \left(
\begin{array}{c}
a_1 \\ a_2
\end{array}\right)_{in} ,
\end{equation}
where $a_1$, $a_2$ are the annihilation operators of modes 1
and 2 respectively, $\omega$ describes
the reflectivity and
transmittance of the beam splitter, and
$\phi$ describes the effect of
the phase shifter (in the factorization method given by M. Reck 
{\it{et
al.}} \cite{reck}, the phase shifters described by $\phi$ should be 
placed
at the input ports).
Therefore, we can use appropriate
beam splitters, phase shifters and a mirror
to construct the desired eight-port.

Finally, photon detection is performed at the four output ports.
We can design the total transformation in such a way that if the 
photon
is detected at the first output port, we claim with certainty that
the initial state was $|\psi_1 \rangle$, if the photon is detected at
the second or the third output port, we claim with certainty that
the initial state was either $|\psi_2 \rangle$ or $|\psi_3 \rangle$,
but we do not know which of these two states it was. If the photon
is detected at the fourth output port, we obtain no information
about the input state.

\begin{figure}[ht]
\epsfig{file=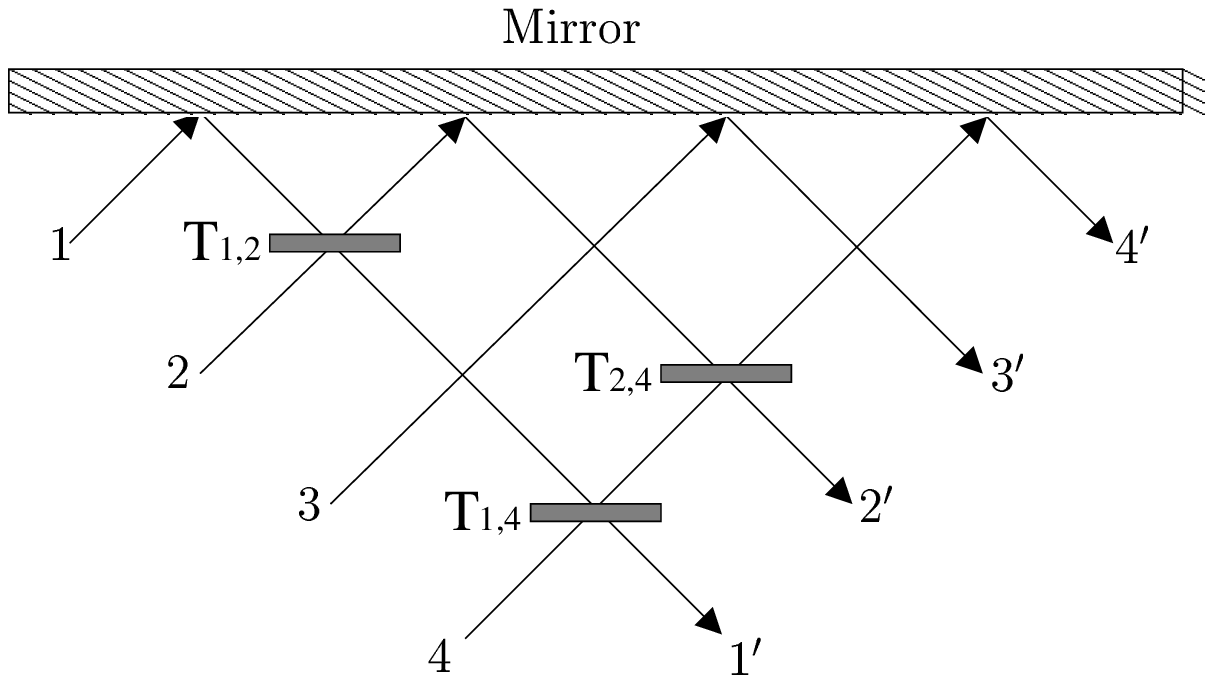, height=5cm}
\caption{The eight-port described by Eq.\ (\ref{m4ex1}) can be
constructed from three beam splitters and a mirror.}
\label{etport2}
\end{figure}

We shall now consider two examples. The first is more general than
the second, but the second has the advantage that it is simple and
the eight-port that it requires consists of only two $50-50$ beam
splitters. In the first example, all of the input vectors have the
same overlap, which is given by $s$, and we shall consider the case
$0<s\leq 1/\sqrt{2}$. The optimal failure probabilities for this
case are given in Eq.\ (\ref{symSolution1}).
For the input vectors we shall take
\begin{eqnarray} 
\label{ex1in} 
|\psi_{1}\rangle_{in} &=&\left(
\begin{array}{c}\frac{1}{\sqrt{3}}(1+2s)^{1/2}\\
\sqrt{\frac{2}{3}}(1-s)^{1/2} \\ 0 \\ 0 \end{array}\right) , 
\nonumber \\
|\psi_{2}\rangle_{in} &=&\left( \begin{array}{c}
\frac{1}{\sqrt{3}}(1+2s)^{1/2}\\
-\frac{1}{\sqrt{6}}(1-s)^{1/2} \\ \frac{1}{\sqrt{2}} (1-s)^{1/2} \\ 0
\end{array}\right) ,
\nonumber \\ 
|\psi_{3}\rangle_{in} &=&\left( \begin{array}{c}
\frac{1}{\sqrt{3}}(1+2s)^{1/2} \\
-\frac{1}{\sqrt{6}} (1-s)^{1/2} \\ -\frac{1}{\sqrt{2}} (1-s)^{1/2} \\ 
0
\end{array}\right) .
\end{eqnarray} 
The output vectors, $|\psi_{i}\rangle_{out} =|\psi^{\prime}_{i} 
\rangle
+|\phi_{i}\rangle$,
can be computed by the method outlined above. Doing so gives us
\begin{eqnarray} 
|\psi_{1}\rangle_{out} &=&\left( \begin{array}{c}
0 \\ (1-\sqrt{2}s)^{1/2} \\ 0 \\ (s\sqrt{2})^{1/2}
\end{array}\right) , \nonumber \\
|\psi_{2}\rangle_{out} &=&\left(\begin{array}{c}
((1+s-s\sqrt{2})/2)^{1/2} \\ 0 \\ ((1-s)/2)^{1/2} \\
(s/\sqrt{2})^{1/2} \end{array} \right) , \nonumber \\
|\psi_{3}\rangle_{out}&=&\left(\begin{array}{c}
((1+s-s\sqrt{2})/2)^{1/2} \\ 0
\\ -((1-s)/2)^{1/2} \\ (s/\sqrt{2})^{1/2} \end{array} \right) .
\label{output1}
\end{eqnarray} 
Our next step is to determine the transformation, $U$, that describes
the eight-port, or, more specifically, the matrix $M(4)$ that 
describes
its action in the one-photon subspace. It must satisfy
$|\psi_{i}\rangle_{out} = U|\psi\rangle_{in}$,
and, in addition, it must map the vector that is orthogonal to all
three input vectors, onto the vector that is orthogonal to all three
output vectors, 
\begin{equation} 
\frac{1}{A}\left(\begin{array}{c} -(s\sqrt{2})^{1/2}B \\
-(s\sqrt{2})^{1/2}C \\ 0 \\
BC \end{array}\right) =
M(4)\left(\begin{array}{c} 0 \\ 0 \\ 0 \\ 1 \end{array}\right) ,
\end{equation} 
where 
\begin{eqnarray} A&=&[(1-s)(1+2s)]^{1/2} , \nonumber \\
B&=&(1-s\sqrt{2})^{1/2} ,
\nonumber \\ C&=&(1+s-s\sqrt{2})^{1/2} .
\end{eqnarray} 
These equations determine M(4) and it is given by $M(4)=$
\begin{equation} \label{m4ex1}
\left(\begin{array}{cccc}\sqrt{\frac{2}{3}}\frac{C}{\sqrt{1+2s}} &
-\frac{C}{\sqrt{3(1-s)}} & 0 & -\frac{B}{A}(s\sqrt{2})^{1/2} \\
\frac{B}{\sqrt{3(1+2s)}} & \sqrt{\frac{2}{3}}\frac{B}{\sqrt{1-s}}
& 0 &
-\frac{C}{A}(s\sqrt{2})^{1/2} \\ 0 & 0 & 1 & 0 \\ \frac
{(\sqrt{2}+1)(s\sqrt{2})^{1/2}}{\sqrt{3(1+2s)}} & \frac{(\sqrt{2}-1)
(s\sqrt{2})^{1/2}}{\sqrt{3(1-s)}} & 0 & \frac{BC}{A}
\end{array}\right) .
\end{equation} 
This matrix can be expressed as the product of three matrixes each of
which corresponds to a beam splitter. In particular, we have that
\begin{equation} 
\label{decomp1} 
M(4)=T_{2,4}T_{1,4}T_{1,2}
\end{equation} 
where the matrix $T_{p,q}$ represents the action of a beam splitter
that mixes only modes $p$ and $q$. The $4\times 4$ matrix for
$T_{p,q}$ can be obtained from that of a $4\times 4$ identity matrix,
$I$, by replacing the matrix elements $I_{pp}$ and $I_{qq}$ by the
transmissivity of the beam splitter, $t$, replacing $I_{pq}$ by the
reflectivity, $r$, and replacing $I_{qp}$ by $-r$. The
transmissivities and reflectivities for beam splitters in Eq.
(\ref{decomp1}) are
\begin{equation} 
\begin{array}{lll} 
T_{2,4}:{}\ & t=B & r=-(s\sqrt{2})^{1/2} \\ T_{1,4}:{}\ &
t=\frac{C}{A}& 
r=-(s\sqrt{2})^{1/2}\frac{B}{A} \\ T_{1,2}:{}\ &
t=\sqrt{\frac{2(1-s)}{3}} &
r=-\sqrt{\frac{1+2s}{3}} . \end{array}
\end{equation} 
This constitutes a complete description of the optical network
that optimally discriminates between $|\psi_{1}\rangle_{in}$
and $\{ |\psi_{2}\rangle_{in}, |\psi_{3}\rangle_{in}\}$, where
these input states are given in Eq.\ (\ref{ex1in}), and it is
shown schematically in Figure\ \ref{etport2}.

An especially simple network will suffice for our second example.
The input vectors are
\begin{eqnarray}
|\psi_{1}\rangle_{in}&=&\left(\begin{array}{c} \sqrt{2/3} \\ 0 \\
1/\sqrt{3} \\ 0 \end{array}\right) , \nonumber \\
|\psi_{2}\rangle_{in}&=&\left(\begin{array}{c} 0 \\
1/\sqrt{3} \\ \sqrt{2/3} \\ 0 \end{array}\right) , \nonumber \\
|\psi_{3}\rangle_{in}&=&\left(\begin{array}{c} 0 \\
-1/\sqrt{3} \\ \sqrt{2/3} \\ 0 \end{array}\right) .
\end{eqnarray}
These input states have the property that
\begin{eqnarray}
\,_{in}\langle\psi_{1}|\psi_{2}\rangle_{in}&=&\,_{in}\langle\psi_{1}
|\psi_{3}\rangle_{in} =\frac{\sqrt{2}}{3} , \nonumber \\
\,_{in}\langle\psi_{2}|\psi_{3}\rangle_{in}&=&\frac{1}{3} .
\end{eqnarray}
The optimal failure probabilities are found to be $q_{1}=2/3$ and
$q_{2}=q_{3}=1/3$. Using Eqs.\ (\ref{quneven}) and (\ref{Quneven})
this gives
\begin{eqnarray}
Q=\frac{4}{9},
\end{eqnarray}
for the minimum average failure probability of this kind of
generalized measurement. This is to be compared to $5/9$,
the average failure probability of a von Neuman type projective
measurement.

The output vectors, $|\psi_{i}\rangle_{out} =|\psi^{\prime}_{i}
\rangle +|\phi_{i}\rangle$, can again be computed by the method
outlined previously. Doing so gives us
\begin{eqnarray}
|\psi_{1}\rangle_{out} &=&\left( \begin{array}{c}
1/\sqrt{3} \\ 0 \\ 0 \\ \sqrt{2/3} \end{array}\right) ,
\nonumber \\
|\psi_{2}\rangle_{out} &=&\left(\begin{array}{c}
0 \\ 1/\sqrt{3} \\ 1/\sqrt{3} \\ 1/\sqrt{3} \end{array} \right) ,
\nonumber \\
|\psi_{3}\rangle_{out} &=&\left(\begin{array}{c}
0 \\ -1/\sqrt{3} \\ 1/\sqrt{3} \\ (1/\sqrt{3} \end{array} \right) .
\label{output2}
\end{eqnarray}

The matrix $M(4)$ can be chosen to be
\begin{equation}
M(4)=\left( \begin{array}{cccc} 1/\sqrt{2} & 0 & 0 & -1/\sqrt{2} \\
0 & 1 & 0 & 0 \\ -1/2 & 0 & 1/\sqrt{2} & -1/2 \\ 1/2 &
0 & 1/\sqrt{2} & 1/2 \end{array}\right) ,
\end{equation}
and it can be expressed as
\begin{equation}
M(4)=T_{3,4}T_{1,4} .
\end{equation}
In this case, both $T_{1,4}$ and $T_{3,4}$ represent $50-50$
beam splitters, and they are given explicitly by
\begin{eqnarray}
T_{1,4}:{}\ & t=\frac{1}{\sqrt{2}} {}\ & r=-\frac{1}{\sqrt{2}}
\nonumber \\
T_{3,4}:{}\ & t=\frac{1}{\sqrt{2}} {}\ & r=-\frac{1}{\sqrt{2}} .
\end{eqnarray}
This last example constitutes what is probably the simplest choice
of the set of parameters for a possible experimental realization.

\section{Conclusions}
The usual problem considered when trying to unambiguously
discriminate among quantum states is to correctly identify
which state a given system is in when one knows the set of
possible states in which it can be prepared. Here we have
considered a different problem. The set of possible states
is divided into two subsets, and we only want to know to which
subset the quantum state of our given system belongs. As
this is a less ambitious task than actually identifying the
state, we expect that our probability to be successful will
be greater for attaining this more limited goal.

We considered the simplest instance of this problem, the situation
in which we are trying to discriminate between a set containing
one quantum state and another containing two. A method for
finding the optimal strategy for discriminating between these
two sets was presented, and analytical solutions for particular
cases were given. In addition, we have shown that if the
quantum states are single-photon states, where the photon can
be split among several modes, the optimal discrimination strategy
can be implemented by using a linear optical network.

These ideas can be extended in a number of different ways. One
possibility is to consider the situation in which one is given
$N$ qubits, each of which is in either the state $|\psi_{1}\rangle$
of $|\psi_{2}\rangle$, where these states are not orthogonal.
What we would like to know is how many of the qubits are in the
state $|\psi_{1}\rangle$. In order to phrase this problem in a way
that makes its connection to the problems considered in this paper
clear, we note that the total set of possible states
for this problem consists of $2^N$ states (the states are strings
of $N$ qubits), and this can be
divided up into the subsets $S_{n}$, where the members of $S_{n}$
are sequences of $N$ qubits in which $n$ are in the state
$|\psi_{1}\rangle$. For a given sequence of qubits, our problem is
to determine to which of the sets $S_{n}$ it belongs. Another
possibility is to use these methods to compare strings of qubits in
order to find out if they are identical or not. Again, suppose that
we have strings of $N$ qubits in which each qubit is in one of the
two non-orthogonal states, $|\psi_{1}\rangle$ or $|\psi_{2}\rangle$.
We are 
given two of these strings and want to know if they are the same or
not. In this case, our set of possible states consists of pairs of
strings, and hence has $2^{2N}$ members. This is divided into two
subsets, the first, $S_{equal}$, consisting of pairs of identical
$N$-qubit strings ($2^{N}$ members), and its complement,
$\overline{S}_{equal}$, consisting of everything else. Our task,
when given two sequences of $N$ qubits, is to decide if they are in
$S_{equal}$ or in $\overline{S}_{equal}$ \cite{barnett}. More
detailed consideration of these problems remains for future research.

\begin{acknowledgments}
We want to thank S. Barnett, A. Chefles, U. Herzog and I. Jex for
helpful discussions of various aspects of this problem. This research
was supported by the Office of Naval Research
(Grant Number: N00014-92J-1233), by the National Science
Foundation (Grant Number: PHY-9970507), by the
Hungarian Science Research Fund (OTKA, Grant Number: T 030671)
and by a grant from PSC-CUNY as well as by a CUNY collaborative grant.
\end{acknowledgments}

\appendix*
\section{Derivation of the optimal solution via the method of
Lagrange multipliers}

In this section, we shall show that by using the method of
Lagrange multipliers, we can derive the conclusions contained in
Eqs.\ (\ref{result1})-(\ref{Q3}) rigorously, starting from
the fact that for optimal discrimination, the vectors
$|\phi_i \rangle$ must be linearly dependent. To express this 
statement in a compact form we define the positive semidefinite 
matrix $C$, where $C_{ij}=\langle \phi_i|\phi_j \rangle$. Then, in 
general, if $|\phi_i \rangle$ ($i=1,\ldots, n)$ are linearly 
dependent, the determinant of matrix $C$ must vanish,  
$\Delta = \det (C)=0$ \cite{sun2}. With the help of Eqs.\ (\ref{q_i}) 
and \ (\ref{phi_1213}), we can eliminate two of the three overlaps 
from the matrix $C$ and obtain explicitly
\begin{eqnarray}
\label{delta}
\Delta &=& \left|
\begin{array}{ccc}
q_1 & O_{12} & O_{13}
\\ 
O_{12}^{\ast} & q_2 & r e^{i \theta}
\\
O_{13}^{\ast} & r e^{-i \theta} & q_3
\end{array} \right| \nonumber \\
&=&q_{1}q_{2}q_{3} - r^{2}q_{1} - |O_{13}|^{2}q_{2} - |O_{12}|^{2}q_{3} 
\nonumber \\ 
&&+ 2|O_{12}||O_{13}|r\cos(\theta - \alpha) = 0 .
\end{eqnarray}
Here $O_{ij}$ again denotes $\langle \psi_i|\psi_j \rangle$, 
$r e^{i \theta}=\langle \phi_2|\phi_3 \rangle$ is the remaining 
overlap where $r$ and $\theta$ are to be determined from the conditions 
for optimum, and $\alpha = - \arg (O_{12} 
O_{13}^{\ast})$. Since $C$ is positive semidefinite,
all the diagonal subdeterminants of $\Delta$ must be non-negative.

We now wish to minimize the average probability of failure $Q$, Eq.\ 
(\ref{Psf}), subject to the constraint in
Eq. (\ref{delta}). This can be done by minimizing the quantity
\begin{equation}
\tilde{Q} = \sum_{i}^{3} \eta_i q_{i} +\lambda \Delta ,
\end{equation}
where $\lambda$ is a Lagrange multiplier.
The conditions for minimum with respect to $r$ and $\theta$, 
$\partial \tilde{Q}/\partial r =0$ and $\partial \tilde{Q}/\partial 
\theta =0$, lead immediately to
\begin{eqnarray}
\label{e7}
|O_{12}||O_{13}|\cos(\theta -\alpha)- q_1r = 0, \\
\label{e8}
r |O_{12}||O_{13}| \sin(\theta - \alpha) = 0.
\end{eqnarray}
The solutions of these equations, corresponding to the minimum of 
$Q$, are
\begin{equation}
    \label{theta}
    \theta = \alpha ,
\end{equation}
and
\begin{equation}
\label{q1r}
q_1 r = |O_{12}||O_{13}| .
\end{equation}
Next, we perform the optimization with respect to the remaining 
variables. Notice that the derivative of $\tilde{Q}$ with respect to 
$\lambda$ returns Eq.\ (\ref{delta}). Therefore, we use the optimal 
values of $r$ and $\theta$ in Eq. (\ref{delta}) and in the 
conditions for minimum with respect to the failure probabilities, 
$\partial \tilde{Q}/\partial q_{i}=0$ for $i=1,2,3$. After some 
algebra we obtain the following set of equations
\begin{eqnarray}
    \label{delta2}
    q_{1}\Delta &=& \Delta_{12} \Delta_{13} = 0 , \\
    \label{e4}
    q_{1}^{2}\frac{\partial \tilde{Q}}{\partial q_1} &=& \eta_1 
    q_{1}^{2} + \lambda (\Delta_{12}\Delta_{13} \nonumber \\
    &&+|O_{12}|^{2}\Delta_{13} + |O_{13}|^{2}\Delta_{12}) = 0, \\
    \label{e5}
    \frac{\partial \tilde{Q}}{\partial q_2} &=& \eta_2+ \lambda 
    \Delta_{13} = 0, \\
    \label{e6}
    \frac{\partial \tilde{Q}}{\partial q_3} &=& \eta_3+ \lambda 
    \Delta_{12} = 0,
\end{eqnarray}
where $\Delta_{12}$ and $\Delta_{13}$ are the diagonal subdeterminants 
of $\Delta$,
\begin{eqnarray}
    \label{delta12}
    \Delta_{12}=q_1 q_2-|O_{12}|^{2} , \\
    \label{delta13}
    \Delta_{13}=q_{1} q_{3} - |O_{13}|^{2}.
\end{eqnarray}

We now have four variables $q_1, q_2, q_3$, and $\lambda$,
and four equations, Eqs. (\ref{delta2})--(\ref{e6}), to find them.
Eq.\ (\ref{delta2}) tells us that at least one of the diagonal 
subdeterminants vanishes. With no loss of generality we can assume this 
to be $\Delta_{12}=0$. Comparing this to Eq. (\ref{e6}) we see that 
$\lambda$ must be singular. The singularity, however, is tractable since 
the same equation tells us that the product $\lambda\Delta_{12}$ is 
finite. Then it follows from the singular behavior of $\lambda$ and Eq. 
(\ref{e5}) that the other diagonal subdeterminant also vanishes, 
$\Delta_{13} = 0$, but the product $\lambda\Delta_{12}$ also remains 
finite. Using these finite values from Eqs. (\ref{e5})--(\ref{e6}) in 
Eq. (\ref{e4}), we can summarize our findings as follows
\begin{equation}
\label{TwoDeltas2} 
\Delta_{12} = \Delta_{13} = 0,
\end{equation}
which is just equation\ (\ref{TwoDeltas}), and
\begin{equation}
\label{delta6}
\eta_1q_1^2 -\eta_2|O_{12}|^2-\eta_3|O_{13}|^2 
+\lambda \Delta_{12} \Delta_{13} = 0 .
\end{equation}
Multiplying Eq. (\ref{e5}) by $\Delta_{12}$ (or Eq. (\ref{e6}) 
by $\Delta_{13}$) and taking into account Eq. (\ref{TwoDeltas2}) gives 
that the singularity in $\lambda$ is such that 
$\lambda\Delta_{12}\Delta_{13}=0$. Using this in Eq. (\ref{delta6}) 
we finally obtain   
\begin{eqnarray}
\label{solution}
\eta_{1}q_1^2 - \eta_{2}|O_{12}|^2 -\eta_{3}|O_{13}|^2 = 0.
\end{eqnarray} 
This is the solution found in Section II, Eq.\ (\ref{q1opt}), and the 
rest of Section II follows from here and Eq. (\ref{TwoDeltas2}).

For the sake of completeness we also give the expression for 
$1/\lambda$,
\begin{equation}
    \frac{1}{\lambda} = - 
    \sqrt{\frac{\Delta_{12}\Delta_{13}}{\eta_{2}\eta_{3}}} ,
\end{equation}
which exhibits no singularity. In fact, $1/\lambda=0$ when 
$\Delta_{12}=\Delta_{13}=0$, as expected.
Finally, let us note that Eq.\ (\ref{TwoDeltas2}), which is identical
to Eq.\ (\ref{TwoDeltas}), implies that all of the failure vectors,
$|\phi_i \rangle$, are parallel to each other, i. e. they
lie in a space, $\mathcal{A}$, of dimension one.

\end{document}